\documentclass[twocolumn,prb,nofootinbib,nobibnotes]{revtex4}

\usepackage{graphicx}
\usepackage{dcolumn}
\usepackage{bm}
\usepackage{amsmath}
\usepackage{amssymb}

\newcommand{\var}{\operatorname{var}}

\begin{document}

\title{Slow Vibrations in Transport through Molecules}

\author{Tero T. Heikkil{\"a}}
\altaffiliation{Corresponding author. Present address: Low
Temperature Laboratory, Helsinki University of Technology, P. O.
Box 3500, FIN-02015 TKK, Finland} \email{Tero.T.Heikkila@hut.fi}

\author{Wolfgang Belzig}
\affiliation{Institut f{\"u}r Physik, Universit{\"a}t Basel,
Klingelbergstr.~82, CH-4056 Basel, Switzerland }

\date{\today}

\begin{abstract}
We show how one can measure the signal from slow jumps of a single
molecule between metastable positions using a setup where the
molecule is fixed to one lead, and one of the coupling strengths
is controlled externally. Such a measurement yields information
about slow processes deforming the molecule in times much longer
than the characteristic time scales for the electron transport
process.
\end{abstract}

\maketitle

One of the key ideas in studies of electron transport through
single molecules is the aim to relate the properties of the
studied microscopic molecule to the current flowing through it.
Then measuring this current will yield information about the
molecule. There are many interesting transport phenomena, known
from larger structures, e.g., semiconductor quantum dots, that
have been also observed in molecules
\cite{park02,liang02,kubatkin03}. However, perhaps a feature most
specific to the molecular systems is the large signature of the
mechanical vibrations on the transport properties. Such effects
include the electron shuttling \cite{fedorets04} and polaronic
effects \cite{flensberg,galperin05}, e.g., the vibration-assisted
electron tunneling effect, observed through the side peaks in the
differential conductance \cite{stipe98,park00,smit02,pasupathy05}
at positions corresponding to the vibrational frequencies. Another
molecule-specific property can be seen when one is able to vary
the coupling of the molecule to the leads between weak and strong
coupling limits \cite{grueter05}. In this case, one can
quantitatively characterize the different coupling strengths, by
fitting the experimentally measured conductance to a fairly
generic model describing transport through the closest molecular
level(s). Such a model relies on the fact that the molecule is
coupled to the leads only from one side, allowing one to tune the
other coupling over a wide range. From this fit, one then obtains
four molecule-specific parameters corresponding to the two
coupling strengths at given positions, an energy scale describing
the position of the HOMO/LUMO level (whichever is closer) and a
length scale describing the change of the coupling as a function
of the distance. These parameters can then be used as a
fingerprint of that particular molecule.

The typically considered vibrational effects are characteristic of
weak coupling for the electron hopping between the leads and the
molecule, in which case the vibrational frequency scales exceed or
are of the same order as the coupling strength. In such systems,
it is essential to consider the fairly fast and low-amplitude
vibrations inside a single parabolic confining potential around
some long-lived metastable position. However, on a much slower
scale, the molecule may jump between different metastable states
corresponding to different conformations or positions. Our aim is
to discuss in this paper how these jumps may be observed and
characterized.

\begin{figure}[h]
\centering
\includegraphics[width=\columnwidth]{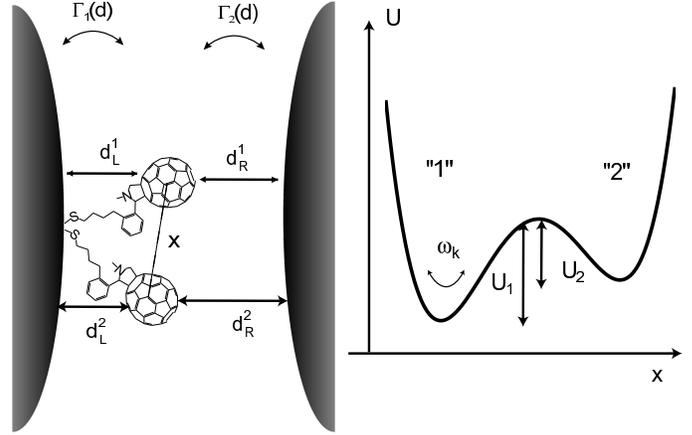}
\caption{Schematic illustration of the considered situation: Left:
A small molecule is connected to the left lead via a linker (this
particular molecule is from Ref.~\onlinecite{grueter05}). Due to
the coupling via a single linker, the molecule-linker-lead system
has multiple metastable configurations, corresponding to different
positions. The distances $d_L^i$ and $d_R^i$ to the two leads are
also indicated. Right: A possible potential profile corresponding
to the position of the molecule, shown in the left figure by the
coordinate $x$. Due to the directional character of the bonding to
the leads, due to the solvent, or due to the particular atomic
arrangement of the left lead, there may be a few metastable
configurations indicated by the potential minima. The hopping
between these configurations depends on the heights $U_1$ and
$U_2$ of the potential barrier. If $U_1 \neq U_2$, the probability
$p_1$ to occupy state $1$ is larger than the corresponding
probability $p_2$ for state 2.} \label{fig:potential}
\end{figure}

Consider a potential energy curve depicted in
Fig.~\ref{fig:potential}. The horizontal axis could quantify
different molecular conformations or average positions. The
vibrations within a single potential well are governed with a
frequency $\omega_k=\sqrt{k/m}$, where $k$ is the spring constant
describing the potential and $m$ is the mass of the molecule.
Brownian motion of the particle within this potential well at the
temperature $T$ will then result into vibrations with amplitude $s
\sim \sqrt{k_B T/k}$. The amplitude of such vibrations is much
smaller than the distance between successive potential minima, and
hence it is at most of the order of a few ${\mathrm \AA}$. We can
use this as an estimate for the frequency $\omega_k$. At room
temperature, for the case of a molecule with mass of the order of
1000 $\dots$ 10000 $m_p$, we would thus get $\omega_k \gtrsim 0.1
\dots 1$ THz. These vibrations are damped by a friction force
described by the characteristic rate $\gamma_f=c\eta R/m$, where
$c$ is a constant of order unity depending on the shape of the
molecule ($c=6\pi$ for a spherical molecule), $R$ characterizes
the size of the molecule, and $\eta$ is the viscosity describing
the molecule environment. For a spherical molecule of size $R \sim
1$nm, mass as given above and using the viscosity of water,
$\eta$=1 g/(ms), we would thus get $\gamma_f \sim 10$ THz. Note
that in practice, the effective viscosity of the solvent depends
also on the molecule itself and thus this number should be used as
indicative only. The jumps between the different potential minima
have a much lower rate than the small-scale vibrations. This rate
is described by the Arrhenius law, \cite{haenggi90}
\begin{equation}
\gamma = \nu \exp(-\frac{U_1}{k_B T}),
\end{equation}
where $U_1$ describes the height of the potential barrier (see
Fig.~\ref{fig:potential}), $\nu =
\frac{\omega_c}{\gamma_f}\frac{\omega_k}{2\pi}$ in the overdamped
limit $\gamma_f \gg \omega_c$ and $\nu = \omega_k/(2\pi)$ for
$\omega_k \gg \gamma_f$. Here $\omega_c$ describes the width of
the potential barrier, and is of the same order as $\omega_k$.
With the above estimates for the frequencies $\omega_k$,
$\omega_c$ and $\gamma_f$, the prefactor $\nu$ thus ranges from
GHz to THz. However, the exponential factor makes the jumps
between different minima much less frequent. Assume for example a
potential barrier height of $U \approx 0.5$ eV
\cite{bindingenergynote}. At room temperature, we would then get
$\gamma = \nu \exp(-20)$; ranging between Hz and kHz. This is
close to the characteristic scale in which the measurements on the
molecules are made and indeed such measurements \cite{grueter05}
showed large fluctuations in the measured conductance, clearly
connected to the presence of the molecule.

The distance-dependent linear conductance $G=G_0 T$ through a
single molecular level can be described by the Breit-Wigner
formula \cite{datta},
\begin{equation}
T=\frac{\Gamma_L \Gamma_R}{\epsilon_1^2+(\Gamma_L+\Gamma_R)^2/4}.
\label{eq:bwformula}
\end{equation}
Here $G_0=2e^2/h$, $\epsilon_1$ is the energy of the closest
molecular level to the metal Fermi energy (i.e., LUMO or HOMO,
whichever is closer) assuming it has an appreciable coupling to
the leads, and $\Gamma_L$ and $\Gamma_R$ characterize the coupling
to the left and right leads, respectively. The level $\epsilon_1$
may be degenerate - this degeneracy would only tune the effective
coupling strengths $\Gamma_L$ and $\Gamma_R$ compared to the
non-degenerate case. For simplicity, we neglect interaction
effects. This assumption still captures the essential physics in
the strong-coupling regime where the coupling energy
$\Gamma_L+\Gamma_R$ exceeds the thermal energy and thus describes
the lifetime of the level. Moreover, additional molecular levels
may be considered, but their contribution shows up mostly to
slightly rescale the coupling constants \cite{heikkilaup}.

Consider now what happens if the molecule is connected to one of
the leads, say left, thus fixing the average $\Gamma_L$. Assume
furthermore $\Gamma_L \ll \epsilon_1$. The average coupling to the
right lead depends on the distance $d_R$ between the molecule and
the furthermost atom of this lead through $\Gamma_R = \Gamma_R^*
\exp(-\kappa d_R)$, where $\kappa$ depends on the solvent and on
the molecule/lead materials \cite{grueterup05}. For $\Gamma_R \ll
\epsilon_1$, decreasing $d_R$ will increase the conductance.
However, when the right lead is close enough, $\Gamma_R$ may
exceed the level energy $\epsilon_1$. In this case, the
conductance shows a maximum at
$\Gamma_R=\sqrt{4\epsilon_1^2+\Gamma_L^2}\approx 2\epsilon_1$ and
further decrease of $d_R$ leads to a decrease in the conductance.
This type of a model was employed to explain the observed
conductance-distance curve in Ref.~\onlinecite{grueter05} with a
quantitative agreement between the theory and the measured average
conductance.

Consider now the fluctuation of this conductance, due to the slow
hoppings of the molecule between different average positions. Such
hopping corresponds to a random telegraph noise in a
time-dependent signal. Let us denote the average distance between
the right lead and the molecule by $d$ (average meaning averaging
over the different positions of the molecule corresponding to the
given positions of the leads). Let us furthermore choose the
coupling strengths corresponding to this average position to
$\overline{\Gamma}_L$ and $\overline{\Gamma}_R=\Gamma_R^*
\exp(-\kappa d)$. Then, the fluctuations of the position around
this average position can be characterized by the values $\{\delta
d^i,-c_i\delta d^i\}$, $i$ indexing the different potential
minima, and the two numbers corresponding to the deviations of the
distance to the left and right leads, respectively. In a typical
case, one could expect that if the molecule moves further from the
left lead ($\delta d>0$), it comes closer to the right lead (as in
Fig.~\ref{fig:potential}). This would thus correspond to a
positive $c_i$. However, for certain situations it may be possible
to increase the distance to both leads - this would be described
with a negative $c_i$. With these deviations, the couplings change
to $\Gamma_L^i=\overline{\Gamma}_L \exp(\kappa \delta d^i)$ and
$\Gamma_R^i=\overline{\Gamma}_R \exp(-\kappa c_i \delta d^i)$.
Note that choosing the same $\kappa$ for both $\Gamma_L$ and
$\Gamma_R$ does not mean a lack of generality, as a possible
difference in the two $\kappa$'s can be included to scale $c_i$.

\begin{figure}[h]
\centering
\includegraphics{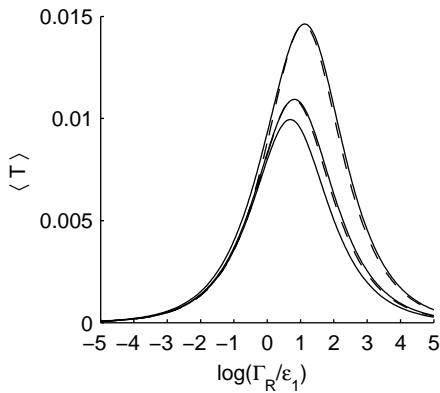}
\caption{Average transmission as a function of coupling $\Gamma_R$
for different amplitudes of variations $\delta d$. The coupling
$\Gamma_R$ can be varied by varying the distance between the
electrodes, and the transmission peak details information about
the molecule and its coupling to the leads. The hopping between
different metastable configurations corresponding to different
$\delta d$ affects the transmission: from top to bottom, $\delta
d=0$ (no hopping), $\delta d=0.5/\kappa$ and $\delta d=1/\kappa$.
We chose $\bar{\Gamma}_L=0.01\epsilon_1$ and $c=0.5$ and describe
hopping between two degenerate configurations (i.e., with equal
probabilities) separated from the average position by $\pm \delta
d$. The dashed lines indicate fits to the Breit-Wigner
transmission with no account of the fluctuations, but with an
increased $\Gamma_L$ and a smaller $\Gamma_R^*$.}
\label{fig:average}
\end{figure}

The transmission averaged over the positions of the molecule is
\begin{equation}
\langle T \rangle = \sum_i p_i T_i \equiv \sum_i p_i
\frac{\Gamma_L^i
\Gamma_R^i}{\epsilon_1^2+(\Gamma_L^i+\Gamma_R^i)^2/4}.
\end{equation}
Here $p_i$ is the probability for the molecule to be in the
position/configuration $i$. Thus, already the average transmission
$\langle T \rangle$ depends on the amplitude $\delta d^i$ of
fluctuations (see Fig.~\ref{fig:average}). However, such a
dependence is difficult to see in $\langle T \rangle$, as a
similar behavior could be observed also without vibrations, but
with a slight rescaling of $\overline{\Gamma}_{L/R}$.

The variance of the transmission values due to these slow
fluctuations is $\var(T)=\langle (T-\langle T \rangle)^2 \rangle$.
Assuming we can neglect the electronic noise (see below) which
also shows up as a temporal variation of the current, this
$\var(T)$ would vanish without the vibrations. In general,
$\var(T)$ depends on $d$, the separation of the leads. However, if
we assume that the positions of the metastable states with respect
to the left lead are independent of $d$, we can separate two
limits in the $d$-dependence. One is the case when $d$ is large,
such that $\Gamma_R \ll \epsilon_1$. If also $\Gamma_L \ll
\epsilon_1$, we can neglect the lifetime of the level (the term
$\Gamma_L+\Gamma_R$ in the denominator of $T_i$). Then
\begin{equation}
T_i \overset{\Gamma \ll 2\epsilon_1}{\rightarrow}
\frac{\Gamma_L^i\Gamma_R^i}{\epsilon_1^2}=\frac{\overline{\Gamma}_L\overline{\Gamma}_R}{\epsilon_1^2}
\exp(\kappa(1-c_i)\delta d^i).
\end{equation}
Thus, the transmission probability for each $i$ can be written in
a form $T_i=\overline{T} T_{f1}^i$, where $\overline{T}$ is
independent of the random hoppings, but depends on the position
$d$, and $T_{f1}^i$ depends on the hoppings, but not on the
position $d$. In this case, we may express the relative variance
as
\begin{equation}\label{eq:relvar}
\begin{split}
&\sigma_T^2 \equiv \frac{\var(T)}{\langle T
\rangle^2}=\frac{\var(T_{f1}^i)}{\langle T_{f1}^i \rangle^2}\\&=
\frac{\sum_i p_i \left\{\exp[\kappa \delta d_i (1-c_i)]-\sum_j p_j
\exp[\kappa \delta d_j (1-c_j)]\right\}^2}{\left\{\sum_j p_j
\exp[\kappa \delta d_j (1-c_j)]\right\}^2}. \end{split}
\end{equation}
Thus, this quantity no longer depends on the exact value of $d$,
as long as $\overline{\Gamma}_L+\overline{\Gamma}_R \ll
2\epsilon_1$.

The same happens in the opposite limit, $\overline{\Gamma}_R \gg
2\epsilon_1$. In this case, we can neglect all other terms but
$\Gamma_R$ from the denominator of the transmission and
\begin{equation}
T_i \overset{\Gamma_R \gg 2\epsilon_1}\rightarrow
\frac{4\Gamma_L^i}{\Gamma_R^i}=\frac{\overline{\Gamma}_L}{\overline{\Gamma}_R}
\exp(\kappa(1+c_i)\delta d^i)\equiv \bar{T}T_{f2}^i.
\end{equation}
The relative fluctuations $\sigma_T$ again follow
Eq.~\eqref{eq:relvar}, with the only exception that the sign of
each $c_i$ is reversed.

\begin{figure}[h]
 \centering
\includegraphics{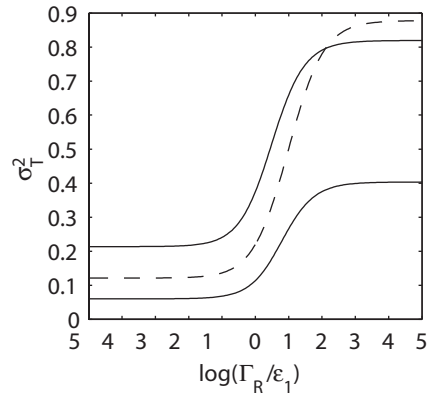}
\caption{Relative variance in the measured conductance curves,
observed as time-dependent fluctuations. The two solid lines are
for hopping between two degenerate configurations with $\kappa
\delta d=0.5$ (bottom) and $\kappa \delta d=1$ (top), and the
dashed curve represents the case with four degenerate
configurations, at $\delta d =\pm 0.5/\kappa$ and $\delta d = \pm
1/\kappa$. We chose $c=0.5$ for each curve.}\label{fig:relvar}
\end{figure}

If the sign of $c_i$ is predominantly positive, $\sigma_T$ in the
case $\Gamma_R \gg 2\epsilon_1$ will be larger than in the case
$\Gamma \ll 2\epsilon_1$ and vice versa for a predominantly
negative $c_i$. Thus, we can sketch the rough behavior of
$\sigma_T$ as a function of the distance $d$ (assuming $c>0$): At
first, when the leads are far apart, $\sigma_T$ stays mostly
constant. When $\overline{\Gamma}_R$ becomes of the order of
$2\epsilon_1$, $\sigma_T$ starts to increase with $d$, until
saturating into another constant value at $\overline{\Gamma}_R \gg
2\epsilon_1$ (see Fig.~\ref{fig:relvar}). Such a behavior holds as
long as the metastable positions of the molecule are unaffected by
the right lead. The latter type of a mechanical effect would show
up also in the average conductance curves if the right lead
changes the potential landscape seen by the molecule. This was
probably observed in Ref.~\onlinecite{grueter05}, but only when
$\bar{\Gamma}_R$ was already much larger than $\epsilon_1$.

To explore this behavior explicitly, let us consider a simple
two-position model with the positions $\{\delta d/2,-c\delta
d/2\}$ and $\{-\delta d/2,c\delta d/2\}$. In this case, we get a
fairly simple expression for $\sigma_T$ in the limit
$\overline{\Gamma}_L \ll \epsilon_1$,
\begin{equation}
\sigma_T=\left|\frac{4 \epsilon_1^2 \sinh(\kappa (1-c) \delta d/4)
+ \overline{\Gamma}_R^2 \sinh(\kappa(1+c) \delta d/4)}{4
\epsilon_1^2 \cosh(\kappa (1-c) \delta d/4)+\overline{\Gamma}_R^2
\cosh(\kappa (1+c)\delta d /4)}\right|.
\end{equation}
In the limit $\Gamma_R \ll 2\epsilon_1$, this gives
\begin{equation}
\sigma_T \overset{\Gamma_R \ll 2\epsilon_1}{\rightarrow}
|\tanh(\kappa (1-c) \delta d/4)|,
\end{equation}
and in the opposite limit $\Gamma_R \gg 2\epsilon_1$,
\begin{equation}
\sigma_T \overset{\Gamma_R \gg 2\epsilon_1}{\rightarrow}
|\tanh(\kappa (1+c) \delta d/4)|.
\end{equation}
These limits follow the qualitative discussion above.

Apart from hopping between different positions, in some cases one
may also envisage the molecule to hop between different
conformations on the slow time scales. Such a change in the
conformation in general may lead to a change both in the energy
level $\epsilon_1$ and in the coupling strengths $\Gamma_{L/R}$.
This behavior can be illustrated by considering the simplest case
of hopping between two conformations corresponding to the energies
$\epsilon_1 \pm \delta \epsilon/2$ and couplings $\Gamma_L \pm
\delta \Gamma_L/2$, $\Gamma_R \pm \delta \Gamma_R/2$. The relative
variance $\sigma_T$ now depends on the relative magnitude of these
changes: if $\delta \Gamma_{L/R}/(\Gamma_L+\Gamma_R) \gg \delta
\epsilon/\epsilon_1$, the behavior is analogous to that discussed
above. In the opposite limit of large $\delta \epsilon$, the
relative variance of the conductance values
$\sigma_T=\var(T)/\langle T \rangle^2$ is given by
\begin{equation}
\sigma_T = \frac{16 \delta \epsilon^2 \epsilon_1^2}{(\delta
\epsilon^2 + (\Gamma_L+\Gamma_R)^2 + 4\epsilon_1^2)^2}.
\end{equation}
Thus, the relative variance is largest when the couplings are much
smaller than the level energies, and it decreases as either of the
couplings is increased. A similar conclusion can be drawn for the
general case with many different conformations, along the same
arguments as above.

There are a few experimental constraints for the observation of
the predicted behavior in the fluctuations, characterized by the
different time scales in the problem. An easily satisfied
condition is that the measurement time $\tau_m$ should exceed the
time scales $1/\omega_k$, $\hbar/\Gamma_{L/R}$, $\tau_e=e/\langle
I \rangle$ characterizing the individual charge transport
processes (typically between ps and ns) by a few orders of
magnitude. Here $\langle I \rangle$ is the average current through
the molecule. In this limit, shot noise yields a contribution
$\sim e/(\tau_m \langle I \rangle)=\tau_e/\tau_m$ to the relative
variance and can hence be neglected. The same applies for the
thermal noise provided that $k_B T/(eV) \tau_e/\tau_m \ll
\sigma_T$, where $V$ is the bias voltage applied over the sample.
Another natural condition is that the time scale $\tau_{\rm var}$
for the variations made in the structure (like changing the
distance between the leads) should be longer than $\tau_m$ and the
time scale $\tau_{\rm hops}=1/\gamma$ for the slow changes in the
configurations. To obtain a relative accuracy $p$ for the measured
variance, one has to measure at least $\sim 1/p^2$ points and
therefore $\tau_{\rm var}/\tau_m > 1/p^2$.

If there are only a few metastable configurations in the problem,
and the time scales for hopping between them is longer than the
measurement time, one may be able to measure the information about
them already by following the telegraph noise in the average
transmission as a function of time. However, for many
configurations, or if at least some of the hopping time scales are
smaller than $\tau_m$, it is better to measure the relative
variance. When $\tau_m$ and $\tau_{\rm hops}$ are well separated,
the measured variance in the signal will be proportional to
\begin{equation}
\var(G)_m=\var(G)_c \frac{\min(\tau_m,\tau_{\rm
hops})}{\max(\tau_m,\tau_{\rm hops})}.
\end{equation}
Here $\var(G)_m$ is the measured variance and $\var(G)_c=G_0^2
\var(T)$ is the variance calculated above. In the case when there
are multiple time scales describing the slow fluctuations, and the
measurement time is between these scales, the measured variance
will be independent of $\tau_m$, characteristic for flicker noise.

Summarizing, in this paper we predict that the different
metastable atomic configurations in molecular junctions have a
considerable effect in the measured conductance, as the time scale
of typical conductance measurements is of the same order as the
time scales for the jumps between the different configurations. We
utilize a simple Breit-Wigner model to illustrate this behavior
and show that such variations lead to a fairly universal behavior
in the relative variance of the measured conductance values as one
of the coupling constants between the molecule and the leads is
controlled.

We thank Christoph Bruder, Michel Calame, Lucia Gr\"uter and
Christian Sch\"onenberger for discussions that motivated this
paper. This work was supported by the Swiss NSF and the NCCR
Nanoscience.


\begin{thebibliography}{19}
\bibitem{park02}
J.~Park, A.~N. Pasupathy, J.~I. Goldsmith, C.~Chang, Y.~Yaish,
P.~J. R.,
  M.~Rinkoski, J.~P. Sethna, H.~D. Abruna, P.~L. McEuen, and D.~C. Ralph,
\newblock Nature {\bf 417}, 722 (2002).
\bibitem{liang02}
W.~Liang, M.~P. Shores, M.~Bockrath, J.~R. Long, and H.~Park,
\newblock Nature {\bf 417}, 725 (2002).
\bibitem{kubatkin03}
S.~Kubatkin, A.~Danilov, M.~Hjort, J.~Cornil, J.-L. Bredas,
N.~Stuhr-Hansen,
  P.~Hedeg\aa rd, and T.~Bjornholm,
\newblock Nature {\bf 425}, 698 (2003).
\bibitem{fedorets04} D. Fedorets, L. Y. Gorelik, R. I. Shekter,
and M. Jonson, Phys. Rev. Lett. {\bf 92}, 166801 (2004).
\bibitem{flensberg} K. Flensberg, Phys. Rev. B {\bf 68},
205323 (2003); S. Braig and K. Flensberg, {\it ibid} {\bf 68},
205324 (2003).
\bibitem{galperin05} M. Galperin, M. A. Ratner, and A.
Nitzan, Nano Lett. {\bf 5}, 125 (2005).
\bibitem{stipe98} B. C. Stipe, M. A. Rezaei, and W. Ho, Science
{\bf 280}, 1732 (1998).
\bibitem{park00}
H.~Park, J.~Park, A.~K.~L. Lim, E.~H. Anderson, A.~P. Alivisatos,
and P.~L.
  McEuen,
\newblock Nature {\bf 407}, 57 (2000).
\bibitem{smit02}
R.~H.~M. Smit, Y.~Noat, C.~Untiedt, N.~D. Lang, M.~C. van Hemert,
and J.~M. van Ruitenbeek, \newblock Nature {\bf 419}, 906 (2002).
\bibitem{pasupathy05} A. N. Pasupathy, {\it et al.}, Nano Lett.
{\bf 5}, 203 (2005).
\bibitem{grueter05} L. Gr\"uter, F. Cheng, T. T. Heikkil\"a, M.
T. Gonzalez, F. Diederich, C. Sch\"onenberger, and M. Calame,
Nanotechnology {\bf 16}, 2143 (2005).
\bibitem{haenggi90} P. H\"anggi, P. Talkner, and M.
Borkovec, Rev. Mod. Phys. {\bf 62}, 251 (1990).
\bibitem{datta}
S.~Datta,
\newblock Nanotechnology {\bf 15}, S433 (2004).
\bibitem{heikkilaup} T. T. Heikkil\"a, C. Sch\"onenberger and W. Belzig,
in preparation.
\bibitem{bindingenergynote} This could arise due to the
energy difference in the different arrangements of the contacting
atoms, for typical energy scales in the case of gold clusters, see
J. Zhao, J. Yang, and J. G. Hou, Phys. Rev. B {\bf 67}, 085404
(2003).
\bibitem{grueterup05} L. Gr\"uter, M. T. Gonzalez, R. Huber, M.
Calame, and C. Sch\"onenberger, Small {\bf 1}, 1067 (2005).
\end{thebibliography}
\end{document}